\documentstyle[epsfig]{mn}

\def\msun{\rm M_{\odot}}
\def\kms{\rm km \, s^{-1}}

\def\simlt{\mathrel{\rlap{\lower 3pt\hbox{$\sim$}}\raise 2.0pt\hbox{$<$}}}

\def\simgt{\mathrel{\rlap{\lower 3pt\hbox{$\sim$}} \raise 2.0pt\hbox{$>$}}}

\def\lsim{\mathrel{\rlap{\lower 3pt\hbox{$\sim$}}\raise 2.0pt\hbox{$<$}}}

\def\gsim{\mathrel{\rlap{\lower 3pt\hbox{$\sim$}} \raise 2.0pt\hbox{$>$}}}

\def\mbulge{M_{\rm Bulge}}

\def\msunpc3{\msun~{\rm {pc^{-3}}}}

\newcommand{\be}{\begin{equation}}

\newcommand{\ee}{\end{equation}}

\def\kms{{\rm\,km\,s^{-1}}}

\begin{document}

\title[Dual massive black holes]{Dual black holes in merger remnants.
  II: spin evolution and gravitational recoil} \author[Dotti et
al.]{M.~ Dotti$^{1,}$$^2$\thanks{e-mail address:
    mdotti@umich.edu}, M.~Volonteri$^1$, A.~Perego$^{3,}$$^4$, M.~Colpi$^4$,
  M.~Ruszkowski$^{1,}$$^5$,
  F.~Haardt$^{2,}$$^6$\\
  $^1$ Department of Astronomy, University of Michigan, Ann Arbor, MI, 48109, USA\\
  $^2$ Dipartimento di Fisica e Matematica, Universit\`a  dell'Insubria, Via Valleggio 11, 22100 Como, Italy\\
  $^3$ Department of Physics, University of Basel, Klingerbergstr. 82,
  4056
  Basel, Switzerland\\
  $^4$ Dipartimento di Fisica G.~Occhialini, Universit\`a degli Studi
  di Milano
  Bicocca, Piazza della Scienza 3, 20126 Milano, Italy\\
  $^5$ The Michigan Center for Theoretical Physics, Ann Arbor, MI, 48109, USA\\
  $^6$ INFN, Sezione di Milano--Bicocca, 20126 Milano, Italy } \maketitle
\vspace {7cm}

\begin{abstract}
Using high resolution hydrodynamical simulations, we explore the spin
evolution of massive dual black holes orbiting inside a circumnuclear
disc, relic of a gas-rich galaxy merger.  The black holes spiral
inwards from initially eccentric co- or counter-rotating coplanar
orbits relative to the disc's rotation, and accrete gas that is
carrying a net angular momentum.  As the black hole mass grows, its
spin changes in strength and direction due to its gravito-magnetic
coupling with the small-scale accretion disc.  We find that the black
hole spins loose memory of their initial orientation, as accretion
torques suffice to align the spins with the angular momentum of their
orbit on a short timescale ($\simlt 1-2$ Myr). A residual off-set in
the spin direction relative to the orbital angular momentum remains,
at the level of $\simlt 10^o$ for the case of a cold disc, and $\simlt
30^o$ for a warmer disc. Alignment in a cooler disc is more effective
due to the higher coherence of the accretion flow near each black hole
that reflects the large-scale coherence of the disc's rotation. If the
massive black holes coalesce preserving the spin directions set after
formation of a Keplerian binary, the relic black hole resulting from
their coalescence receives a relatively small gravitational recoil.  The
distribution of recoil velocities inferred from a simulated sample of
massive black hole binaries has median $ \simlt 70 \kms$, much
smaller than the median resulting from an isotropic distribution of
spins.

\end{abstract}


\section{Introduction}

The massive black holes (MBHs) that we observe today in local
spheroids (Ferrarese \& Ford 2005, and references therein) are
expected to have grown through a series of major accretion episodes in
symbiosis with the growth of their host galaxies. Gas-rich major
mergers may be at the heart of this joint evolution as they may
explain the morphology of the hosts and at the same time account for
the fueling of the underlying MBHs (e.g. Di Matteo, Springel \&
Hernquist 2005, and references therein).  In the currently favored
cold dark matter hierarchical cosmologies, galaxy mergers play indeed
a key role in growing galaxies to their present sizes and the
coalescence of MBHs in binaries is therefore expected to be relatively
common (Menou, Haiman \& Narayanan 2001; Volonteri, Haardt \& Madau
2003).

Following a galaxy major merger, the pair of MBHs first interacts with
stars (Begelman, Blandford \& Rees 1980; Milosavljevic \& Merritt
2001) and gas (Mayer et al. 2007). The pair loses orbital angular
momentum, and it is expected to harden progressively down to subparsec
scales where emission of gravitational radiation drives the MBH
inspiral down to coalescence. Depending on the properties of the
coalescing binary, the pattern of the gravitational wave emission can
be anisotropic, resulting in a non zero recoil velocity (a ``kick'')
of the MBH remnant (Redmount \& Rees 1989).  Several attempts to
compute analytically the strength of the kick have been undertaken
(Peres 1962; Bekenstein 1973; Fitchett 1983; Fitchett \& Detweiler
1984; Redmount \& Rees 1989; Wiseman 1992; Favata, Hughes \& Holz
2004; Blanchet, Qusailah \& Will 2005; Damour \& Gopakumar 2006;
Schnittman \& Buonanno 2007.

Recent numerical simulations of the coalescence of spinning MBHs in
full general relativity have been able to calculate explicitly kick
velocities for a series of different binary configurations. It is
found that three parameters influence the magnitude of the
gravitational recoil of the relic MBH: the binary mass ratio, the
spins, and the mutual orientation of the spins with respect to the
orbital angular momentum. The recoil is largest, up to $4000\,
\rm{km\,s^{-1}}$, for nearly equal mass MBHs with large spins, when
the spin vectors have opposite directions and are in the orbital plane
(Campanelli et al. 2007).  By contrast, recoils of $\lsim 200 \kms$
are imparted to the MBH remnant if the spins of the progenitors prior
coalescence are orthogonal to their orbital plane. 

Purely general relativistic (GR) effects (i.e., spin-orbit and
spin-spin interactions) may produce low recoil configurations, when
the MBH pairing is driven by gravitational wave emission. However,
those GR effects depend strongly on the initial relative orientation
of the MBH spins, and can result in low recoil configurations only for
a small region of parameter space (Schnittman 2004; Herrmann et al.
2009; Lousto et al. 2009).

In gas-rich mergers between galaxies of comparable mass (i.e. major
mergers), close binary MBHs form under the action of dynamical
friction against the gaseous and stellar background (Mayer et al.
2007; Callegari et al. 2009; see Colpi \& Dotti 2009 for a review).
During their inspiral MBHs are surrounded by a dense cocoon of gas
that drives their dynamical decay and provides fuel for the feeding of
the holes (Dotti et al. 2007, 2009).  Since matter carries angular
momentum also the spin vector can change during the accretion process.
The details of the dynamics may have a profound influence on the mass
and spin evolution of the two MBHs, and thus on the recoil velocity of
the MBH resulting from their coalescence, and this is matter of our
concern in this paper.

The spin evolution during the MBH inspiral in a gas rich merger
remnant has many implications .  Spins, prior to coalescence,
influence the extent of the gravitational recoil, and so the retention
of the relic MBH inside its host galaxy.  Accordingly, the spin
distribution of the coalescing binaries, is critical as it determines
the frequency of MBH retention in the host halo (Volonteri \& Rees
2006; Volonteri 2007; Volonteri, Haardt \& Gultekin 2008; Gultekin et
al. in preparation).  The magnitude of the MBH spins and their
orientation relative to the orbit during mergers is also critical in
shaping the stellar density profiles in ellipticals, as the kicked MBH
moving on a return orbit can deposit its excess kinetic energy into
the stellar background, causing the formation of stellar core
(Boylan--Kolchin, Ma \& Quataert 2004; Gualandris \& Merritt 2008). A
recoiling MBH can have an observational signature when moving across
the host galaxy, creating an X-ray tail in the perturbed hot gas
(Devecchi et al. 2009), an off-set active nucleus (Loeb 2007;
Volonteri \& Madau 2008), shocking the inner rim of the accretion disc
(Lippai, Frei \& Haiman 2008; Schnittman \& Krolik 2008), or dragging
a stellar cusp with peculiarly high velocity dispersion (Merritt,
Schnittman \& Komossa 2009). Furthermore, spin orientations have
important implications for gravitational wave astronomy, and for using
gravitational wave measurements to constrain the formation history of
MBHs (Vecchio 2004; Lang \& Hughes 2006; Berti \& Volonteri 2008; Arun
et al. 2009a, 2009b).

Bogdanovic, Reynolds \& Miller (2007) proposed a physical process that
could {\it align} the MBH spins with the orbital angular momentum of
the binary, thus leading to slow recoils for the MBH remnant.
The key process for alignment is the presence of a coherent gas
inflow.  They speculate that accreting gas exerts gravito-magnetic
torques that suffice to align the spins of both the MBHs with the
angular momentum of the large-scale gas flow in which the orbit is
embedded.

Spin--disc alignment due to gravito--magnetic coupling has been
studied by a number of authors in the case of isolated MBHs surrounded
by their own discs (Bardeen \& Petterson 1975; Natarajan \& Pringle
1998; Scheuer \& Feiler 1996, Martin, Pringle \& Tout 2007;
Perego et al. 2009).  Here we attempt to explore for the first time
spin-disc alignment around MBH binaries. We expect low recoils when the
spin--disc coupling is strong, i.e. when: 
\\$\bullet$ The two MBHs accrete $\gsim 1 \%$ of their initial mass
before the coalescence (Natarajan \& Pringle 1998; Natarajan \&
Armitage 1999; Volonteri, Sikora \& Lasota 2007; Perego et al. 2009); 
\\$\bullet$ The accreted gas carries angular momentum in a
preferred direction, flowing onto each MBH along a preferential
plane determined by the distribution of angular momentum of the
gas in the environment of the MBH.\\

The first requirement can be fulfilled if the MBHs pair inside a dense, massive
gaseous nuclear disc (Dotti et al. 2009), such as that predicted to
form in remnants of gas-rich major mergers by Mayer et
al. (2007).  To constrain the second requirement, we analyse a set of
3D Smooth Particle Hydrodynamics (SPH) simulations already discussed
in Dotti et al. (2009). The high resolution of these simulations
enables us to resolve the gravitational sphere of influence of each
MBH during their inspiral inside the circumnuclear disc, and to map
the distribution of angular momentum of the SPH particles in the
MBH vicinity. MBHs are
modeled as {\it sink} particles that can accrete gas particles,
allowing us to constrain the amount of mass accreted onto each MBH,
and the orientation of the MBH spins relative to the angular
momentum of the accreted gas.

The paper is organized as follows: in Section~2 we focus on the SPH
simulations, and describe the semi-analytical algorithm that evolves
the MBH spins; in Section~3 we illustrate our results on the MBH
alignment and our prediction for the recoil velocity of the MBH
remnant; in Section~4 we present our conclusions.
 
\section{Numerical methods}

\subsection{SPH simulations}

We follow the dynamics of MBH pairs in nuclear discs using numerical
simulations run with the N--Body/SPH code {\small GADGET} (Springel,
Yoshida \& White 2001), upgraded to include the accretion physics.
The simulations discussed in this paper are the same as presented in
Dotti et al. (2009). Here we give a short summary of the initial
conditions for the different runs. For a more detailed discussion, we
defer the reader to Dotti et al. (2009).

In our models, two MBHs are placed in the plane of a massive
circumnuclear gaseous disc, embedded in a larger stellar spheroid.
The gaseous disc is modeled with $\approx 2 \times 10^6$ particles,
has a total mass $M_{\rm{Disc}}=10^8 \msun$, and follows a Mestel
surface density profile $\Sigma(R) \propto R^{-1},$ where $R$ is the
radial distance projected into the disc plane.  The outer radius of
the disc is 100 pc.  The massive disc is rotationally supported in $R$
and has a vertical thickness of 8 pc.  The internal energy per unit
mass of the SPH particles scales as $u(R)\propto R^{-2/3}$, where the
value of the temperature at the outer radius of the disc has been set
in order to have the Toomre parameter (Toomre 1964) $Q\geq 3$
everywhere, preventing the fragmentation of the disc (the average
value of $Q$ over the disc surface is $\approx 10$).  Gas is evolved
assuming a polytropic equation of state with index $\gamma=5/3$ or
$\gamma=7/5$.  In the former case, the runs are denoted by ``H'' and
are termed ``hot'' as the temperature is proportional to a higher
power of density than in the latter class of runs (``cold'' cases,
runs denoted by ``C'').  The cold case has been shown to provide a
good approximation to a gas of solar metallicity heated by a starburst
(Spaans \& Silk 2000; Klessen, Spaans, \& Jappsen 2007).  The hot case
instead corresponds to an adiabatic monoatomic gas, as if radiative
cooling were completely suppressed during the merger, for example as a
result of radiative heating after gas accretion onto the MBHs (Mayer
et al. 2007).

 The spheroidal component (bulge) is modeled with $10^5$ collisionless
particles, initially distributed as a Plummer sphere with a total mass
$\mbulge(=6.98 M_{\rm{Disc}})$.  The mass of the bulge within $100$ pc
is five times the mass of the disc, as suggested by Downes \& Solomon
(1998).
 
 The two MBHs ($M_1$ and $M_2$) are equal in mass ($M_{\rm BH}=4\times
10^6\,\msun$). The initial separation of the MBHs is 50 pc.  $M_1$,
called primary for reference, is placed at rest at the centre of the
circumnuclear disc. $M_2$, termed secondary, is moving on an
initially eccentric ($e_0\simeq 0.7$) counterrotating (retrograde MBH,
``R'' runs) or corotating (prograde MBH, ``P'' runs) orbit with
respect to the circumnuclear disc. Given the large masses of the disc
and the bulge, the dynamics of the moving MBH ($M_2$) is unaffected by
the presence of $M_1$ until the MBHs form a gravitationally bound
system.

We allow the gas particles to be accreted onto the MBHs if the
following two criteria are fulfilled:\\ 
 \noindent
$\bullet$ the sum of the kinetic and internal energy of the gas
particle is lower than $b$-times the modulus of its
gravitational energy (all the energies are computed with respect to
each MBH);\\ 
\noindent $\bullet$ the total mass accreted per unit time
onto the MBH every timestep is lower than the accretion rate
corresponding to the Eddington luminosity ($L_{\rm Edd}$) computed
assuming a radiative efficiency ($\epsilon$) of 10\%.\\
 
 \indent The parameter $b$ is a constant that defines the degree 
 at which a  particle is bound to the MBH in order to be accreted. We set
 $b=0.3$. Note that due to the nature of the above criteria, the
 gas particles can accrete onto the MBHs only if the
time-varying  Bondi-Hoyle-Lyttleton radius
is resolved in the simulations. 

Each gas particle accreted by the MBH carries with it angular
momentum.  From the properties of the accreted particles we can
compute, as a function of time, the mass accretion rate and the versor
$\mathbf \hat l_{\rm edge}$, that defines the direction of the
total angular momentum of the accreted particles.

This information can be gathered only by performing very high
resolution simulations. The gravitational softening of the MBHs is 0.1
pc. The gravitational softening of the gas particles is set to the
same value, in order to prevent numerical errors.  This is also the
spatial resolution of the hydrodynamical force in the highest density
regions\footnote{The code computes the density of each SPH particle
  averaging over $N_{\rm neigh}=32$ neighbors.}.  The gravitational
softening of the collisionless particles forming the bulge is 1 pc, in
order to prevent two body interactions between gas particles and
artificially massive stars.  The main input parameters of our
simulations are summarized in Table~1.

\begin{table}
\label{tab:run}
\begin{center}
\caption{Run parameters}
\begin{tabular}{l@{   }c@{   }c@{    }c@{    }c@{   }}  
\hline
\\
run & ~~~prograde ? & ~~~$e_{0}$ & ~~~$\gamma$  \\
\\
\hline
\hline 
\\
HP & ~~~yes &  ~~~0.7  & ~~~5/3, ``hot'' \\
HR & ~~~no  &  ~~~0.7  & ~~~5/3, ``hot''    \\ 
\hline
CP & ~~~yes & ~~~0.7   & ~~~7/5, ``cold'' \\
CR & ~~~no  & ~~~0.7   & ~~~7/5, ``cold''    \\
\hline
\end{tabular}\\
\end{center}
\noindent
\end{table}

\subsection{Semi-analytical Bardeen-Petterson effect}

We use the MBH accretion histories obtained from our SPH
simulations to follow the evolution of each MBH spin vector,
${\mathbf J}_{\rm BH}=(aGM_{\rm BH}^2/c)\hat{{\mathbf J}}_{\rm BH}$,
where $0 \leq a \leq 1$ is the dimensionless spin parameter and
$\hat{\mathbf J}_{\rm BH}$ is the spin versor.  The scheme we adopt
to study the spin evolution is based on the model recently developed
by Perego et al. (2009). Here we summarize this algorithm.

We assume that during any accretion event recorded in our SPH
simulations, the inflowing gas forms a geometrically thin/optically
thick $\alpha$-disc (Shakura \& Sunyaev 1973) on milli-parsec scales
(not resolved in the simulation), and that the outer disc orientation
is defined by the unit vector ${\mathbf l}_{\rm edge}.$ The evolution
of the $\alpha$-disc is related to the radial viscosity $\nu_1$ and
the vertical viscosity $\nu_2$: $\nu_1$ is the standard radial shear
viscosity while $\nu_2$ is the vertical shear viscosity associated to
the diffusion of vertical warps through the disc. The two viscosities
can be described in terms of two different dimensionless viscosity
parameters, $\alpha_1$ and $\alpha_2$, through the relations
$\nu_{1,2}=\alpha_{1,2}Hc_{\rm s}$, where $H$ is the disc vertical
scale height and $c_{\rm s}$ is the sound speed of the gas in the
accretion disc. We further assume $\alpha_2=f_2 / (2 \alpha_1)$, with
$\alpha_1 = 0.1$ and $f_2 = 0.6$ (Lodato \& Pringle 2007). We assume
power law profiles for the two viscosities, $\nu_{1,2} \propto
R^{3/4}$, as in the Shakura \& Sunyaev solution.

As shown by Bardeen \& Petterson (1975), if the orbital angular
momentum of the disc around the MBH is misaligned with respect to the
MBH spin, the coupled action of viscosity and relativistic
Lense-Thirring precession warps the disc in its innermost region forcing
the fluid to rotate in the equatorial plane of the spinning MBH.
The timescale of propagation of the warp is short compared with
the viscous/accretion timescale so that
the deformed disc reaches an equilibrium profile that can be computed
by solving the equation 

\begin{eqnarray} \label{eqn:angular momentum}
\frac{1}{R}\frac{\partial}{\partial R}(R {\mathbf L} v_{\rm R})=
\frac{1}{R}\frac{\partial}{\partial R}\left(\nu_1 \Sigma R^3 \frac{d\Omega}{dR}~ {\mathbf \hat l} \right)+ \nonumber \\
+\frac{1}{R}\frac{\partial}{\partial R}\left(\frac{1}{2}\nu_2 R L \frac{\partial {\mathbf \hat l}}{\partial R} \right) 
+ \frac{2G}{c^2} \frac{{\mathbf J}_{\rm BH} \times {\mathbf L}} {R^3} 
\end{eqnarray}
where $v_R$ is the radial drift velocity, $\Sigma$ is the surface
density, and $\Omega$ is the Keplerian velocity of the gas in the
disc.  $\mathbf{L}$ is the local angular momentum surface density of
the disc, defined by its modulus $L$ and the versor ${\mathbf \hat l}$
that defines its direction.

The boundary conditions to eq.~\ref{eqn:angular momentum} are the
direction of ${\mathbf L}$ at the outer edge ${\mathbf \hat l}_{\rm
edge}$, the mass accretion rate (that fixes the magnitude of
$\Sigma$), and the values of mass and spin of each MBH. All these
values but the MBH spins are directly obtained from the SPH runs.
In particular, the direction of the unit vector ${\mathbf \hat l}_{\rm
edge}$ is computed considering those SPH particles nearing the MBH
gravitational sphere of influence that are accreted according to the
criteria outlined in Section 2.1.

Also the MBH spin changes, not only because of accretion, but in
response to its gravito-magnetic interaction with the disc on a
timescale longer than the time scale of warp propagation (Perego et
al. 2009).  This interaction tends to reduce the degree of
misalignment between the disc and the MBH spin, decreasing with time the
angle between ${\mathbf J}_{\rm BH}$ and ${\mathbf \hat l}_{\rm
edge}$.  The MBH spin evolution is followed by solving for the equation
\begin{equation} \label{eqn:jbh precession-disc}
\frac{d{\mathbf J}_{\rm BH}}{dt} =\dot{M}\Lambda(R_{\rm ISO})\hat
 {\mathbf l}(R_{\rm ISO}) + \frac{4\pi G}{c^2}\int_{\rm
 disc}\frac{{\mathbf L} \times {\mathbf J}_{\rm BH}}{R^2}dR.
 \end{equation}
 The first term in eq.~\ref{eqn:jbh precession-disc} accounts for the
 angular momentum deposited onto the MBH by the accreted particles at
 the innermost stable orbit (ISO), where $\Lambda(R_{\rm ISO})$
 denotes the specific angular momentum at $R_{\rm ISO}$ and $\hat
 {\mathbf l}(R_{\rm ISO})$ the unit vector parallel to ${\mathbf
 J}_{\rm BH}$, describing the warped disc according to the
 Bardeen-Petterson effect.  The second term instead accounts for the
 gravo-magnetic interaction of the MBH spin with the warped disc. It
 modifies only the MBH spin direction (and not its modulus), conserving
 the total angular momentum of the composite (MBH+disc) system (King
 et al. 2005).  The integrand in eq. 2
 peaks at the warp radius ($R_{\rm warp}$) where the disc deformation
 is the largest.\footnote{The exact definition of $R_{\rm warp}$ is
 where the vertical viscous time $R^2/\nu_2$ in the disc is
 comparable to the Lense-Thirring precession time. Because $R_{\rm
 warp}$ and the radius at which the disc is maximally deformed are
 comparable (Perego et al. 2009), we simplify the notation in the paper
 using only $R_{\rm warp}$.}  Eq. 2 incorporates two timescales:
 the accretion time related to the first right-hand term describing
 the $e-$folding increase of the spin modulus, and the shorter
 timescale of MBH spin alignment (Perego et al. 2009)
 \begin{equation}
 \tau_{\rm al} \sim  10^5 a^{5/7} \left(\frac{M_{\rm BH}}{4 \times 10^6 \msun}\right)^{-2/35} f_{\rm Edd}^{-32/35}   {\rm yr},
 \end{equation}
 that will ensure a high degree of MBH-disc gravito-magnetic coupling
 during MBH inspiral, as we will show promptly in Section~3. In Eq~3
 $f_{\rm Edd}$ is the MBH luminosity in units of $L_{\rm Edd}$.
  
We applied iteratively eq. 1 and 2 using inputs from the SPH
simulation that give the values of the mass accretion rate, the MBH
mass and the direction of ${\mathbf \hat l}_{\rm edge}$.  The
algorithm returns, as output, the spin vector, that is, its magnitude
and direction.  At each timestep our code therefore provides the angle
between the spin vector of each MBH and the angular momentum vector of
their relative orbit.

\section{RESULTS}

\begin{figure}
\centerline{\psfig{figure=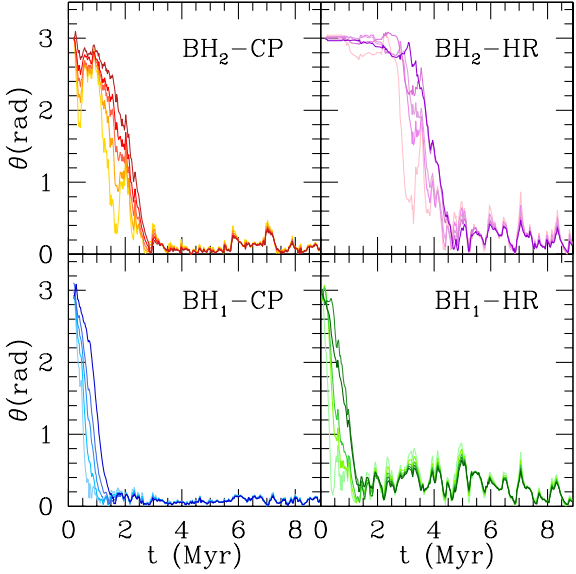,height=8cm}}
\caption{Upper panels: time evolution of the relative angle between
$M_2$ spin and the orbital angular momentum of the MBH pair. Left
(right) panel refers to runs CP (HR). The initial angle is arbitrarily
set to 2.5 radians (close to anti-aligned), and the initial spin
parameter magnitudes varies between 0.2 (lighter colours) to 1 (darker
colours).  Lower panels: same as upper panels for $M_1$.  }
\label{fig:theta}
\end{figure}

Figure~\ref{fig:theta} shows the time evolution of the relative angle
$\theta$ between the spin of each MBH and the orbital angular momentum
of the MBH pair (${\mathbf L}_{\rm pair}=L_{\rm orb}{\mathbf \hat
l}_{\rm pair}$), for two selected runs (CP and HR). The initial
relative angle ($\theta_{\rm i}$) has been arbitrarily set to 2.5
radians (143$^{\circ}$), while $a$ has initially five different values
(0.2, 0.4, 0.6, 0.8, and 1).

There is a common trend in all the runs for both MBHs: MBHs with lower
spins tend to align faster (as shown in Fig.~\ref{fig:theta} for
$t\lsim 2 - 4$ Myr) and are affected by changes in the plane of the
accreting material to a larger extent ($\theta$ changes rapidly with
time and has more pronounced minima/maxima for lower $a$, see again
Fig.~\ref{fig:theta}). As indicated by eq.~3 a smaller spin modulus
implies a shorter alignment time, and this explains the faster
response of the MBH to orient its spin orthogonal to the plane of the
accreted gas. A slowly-spinning MBH induces a weaker warp in the disc:
the warp radius decreases with decreasing $a$ and there the
Lense-Thirring precession time is faster so that the MBH is more
responsive to changes in the orientation of the accreted gas (see
Perego et al. 2009 for details).

The spin evolution depends also on the dynamical properties of the
MBHs and on the thermodynamics of the circumnuclear disc. The effect
of the initial orbital parameters is important during the first
phase of orbital decay of the two MBHs, before they form a binary.
We consider the two MBHs to be bound in a binary if the mass in gas
and stars inside their orbits is lower than the mass of the
binary. This happens when the separation between the two MBHs is
$\approx 5$ pc. The time at which the two MBHs form a binary ($t_{\rm
bin}$) is reported in Table~2 for each run. As described in detail 
in Section~3.1, $M_2$ loses memory of its initial orbital
parameters before binding in a binary. As a consequence, the
properties of accreting gas onto the MBHs after the formation of the
binary are almost independent of the initial dynamical parameters of
the pair.  At this late stage of the orbital evolution, the gas
accretion rate and the coherence of the accretion flows depend mostly
on the thermodynamics of the circumnuclear disc.  Summarizing, for
$t<t_{\rm bin}$ both dynamical and thermodynamical properties affect
spin evolution, while for $t>t_{\rm bin}$ the thermodynamical properties
ultimately determine the final degree of spin alignment.

\begin{table}
\begin{center}
\caption{Third column: MBH binary formation time. Fourth column:
component parallel to ${\mathbf L}_{\rm pair}$ of the angular momentum
of the gas particles accreted after the formation of the binary
($\Delta L_z$), normalized to its modulus ($\Delta L$). Fifth column:
average value of the angle between the MBH spins and $\mathbf L_{\rm
pair}$, after the formation of a binary.}
\begin{tabular}{l@{   }c@{   }c@{   }c@{    }c@{    }}  
\hline
\\
 Run & ~MBH & $t_{\rm bin}$ [Myr] & $\Delta L_z / \Delta L$ & $\theta_{\rm f}$ (rad)\\
\\
\hline
\hline 
CP & ~~~$M_1$ ~~~& 6.5 & ~~~~~$>$99.9\%~~~~~~~ & 0.10\\
CP & ~~~$M_2$ ~~~& 6.5 & ~~~~~$>$99.9\%~~~~~~~ & 0.13\\
CR & ~~~$M_1$ ~~~& 4.5 & ~~~~~$>$99.9\%~~~~~~~ & 0.15\\
CR & ~~~$M_2$ ~~~& 4.5 & ~~~~~$>$99.9\%~~~~~~~ & 0.16\\
HP & ~~~$M_1$ ~~~& 7.5 & ~~~~~~96.3\%~~~~~~~ & 0.25\\
HP & ~~~$M_2$ ~~~& 7.5 & ~~~~~~94.9\%~~~~~~~ & 0.23\\
HR & ~~~$M_1$ ~~~& 4.5 & ~~~~~~81.9\%~~~~~~~ & 0.42\\
HR & ~~~$M_2$ ~~~& 4.5 & ~~~~~~77.9\%~~~~~~~ & 0.32\\
\hline
\end{tabular}\\
\end{center}
\label{tab:acc}
\noindent
\end{table}

\subsection{Effects of dynamics on spin alignment}

We note that for each MBH and every run, $\theta$ initially ($t<4.5$
Myr) decreases with time, but the alignment process is more efficient
for $M_1$ in both simulations. This delay in the alignment of $M_2$ is
related to the orbital evolution of the orbiting MBH. In runs CP and
HP, $M_2$ is initially corotating with the circumnuclear disc on an
eccentric orbit. Because of the eccentricity of the orbit, $M_2$ has a
non-zero relative velocity with respect to its local gas
environment. As a consequence, the accretion rate onto $M_2$ is
initially lower than accretion rate onto $M_1$ (Dotti et
al. 2009). Dynamical friction exerted by the circumnuclear disc onto
the orbiting MBH circularizes the orbit of $M_2$ before the formation
of a binary (Dotti, Colpi \& Haardt 2006a; Dotti et al. 2007), so that
the relative velocity between gas particles and $M_2$ decreases. After
dynamical friction circularized the orbit of $M_2$, the accretion rate
onto $M_2$ increases and becomes comparable to the accretion rate onto
$M_1$ (Dotti et al. 2009). As a consequence the alignment of the spin
of $M_2$ becomes more efficient, and by the time a binary forms,
$\theta$ has similar values for $M_1$ and $M_2$ in the same run.

For initially counterrotating MBHs (runs HR and CR), the effect of the
dynamics onto the spin evolution of $M_2$ is more
pronounced. Dynamical friction drags the orbiting MBH in the direction
of the rotating gas, so that, before the formation of a binary, $M_2$
starts to corotate with respect to the circumnuclear disc (``orbital
angular momentum flip''; Dotti et al. 2009). In the counterrotating
runs the ratio between the accretion rate onto $M_2$ before and after
the angular momentum flip can be $\lsim 0.15$.  As a consequence,
during the first $2-3$ Myrs, when the secondary moves on a retrograde
orbit, $\theta$ does not change significantly (because of the low
accretion rate), while it decreases efficiently only after the orbital
angular momentum flip.

\subsection{Effects of gas thermodynamics on spin alignment}

Alignment occurs over a short time--scale, as indicated by the steep
drop of $\theta$ in Figure~\ref{fig:theta}.  Afterwards, $\theta$
starts to oscillate around an average value, different from run to
run. Numerical noise due to the discrete nature of SPH calculations
does not affect these oscillations. During each oscillation, the MBHs
accrete tens of $N_{\rm neigh}$.  In particular, the average value of
$\theta$ and the amplitude of its oscillations are in general larger
for hot runs (see Figure~\ref{fig:theta}).  We define $\theta_{\rm f}$
as the angle between the MBH spins and $\mathbf L_{\rm pair}$ after
the formation of a binary ($\theta_{\rm f}=\theta(t>t_{\rm bin})$).
This new parameter is of key importance in the following discussion,
since we assume that the distribution of $\theta_{\rm f}$ is
representative of $\theta$ at coalescence. The validity of this
assumption is discussed in Section~3.3.

The last column of Table~2 shows the average value of $\theta_{\rm f}$
of each MBH in every run.  We note that $\theta_{\rm f}$ is lower when
the MBHs are embedded in colder discs. This is due to the properties
of gas close to each MBH. For larger $\gamma$ (hot runs) the
temperature of the gas in the overdense regions around each MBH is
higher, and so is the pressure. As a consequence, the gas structures
around each MBH (and the gas particles accreting onto the MBHs) are
more pressure supported, spherical distributed, and with more
isotropic velocities in runs HP and HR, while gas is more rotationally
supported in runs CP and CR. This effect is quantified in Table~2. In
the fourth column we report the component parallel to ${\mathbf
L}_{\rm pair}$ of the angular momentum of the gas particles accreted
after the formation of the binary ($\Delta L_z$), normalized to its
modulus ($\Delta L$). In cold runs, after the formation of the binary,
streams of gas accreting onto the MBHs are extremely coherent ($\Delta
L_z/\Delta L>99.9\%$). In hot runs the accreting particles have a
larger degree of isotropy, resulting in less coherent accretion
processes ($\Delta L_z/\Delta L \lsim 95\%$) and larger/more variable
$\theta_{\rm f}$.

\begin{figure*}
\includegraphics[width=0.48\textwidth]{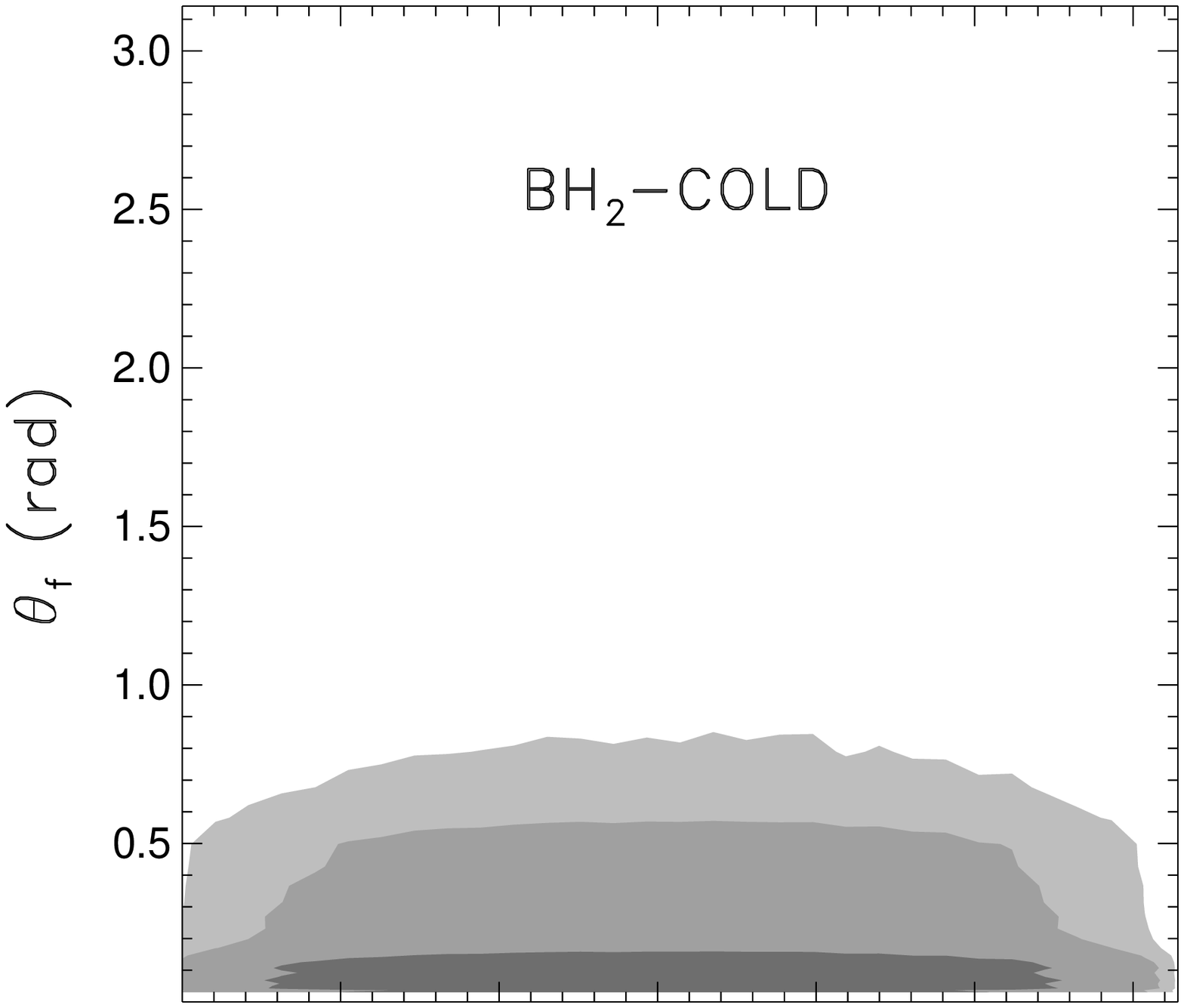}
\includegraphics[width=0.48\textwidth]{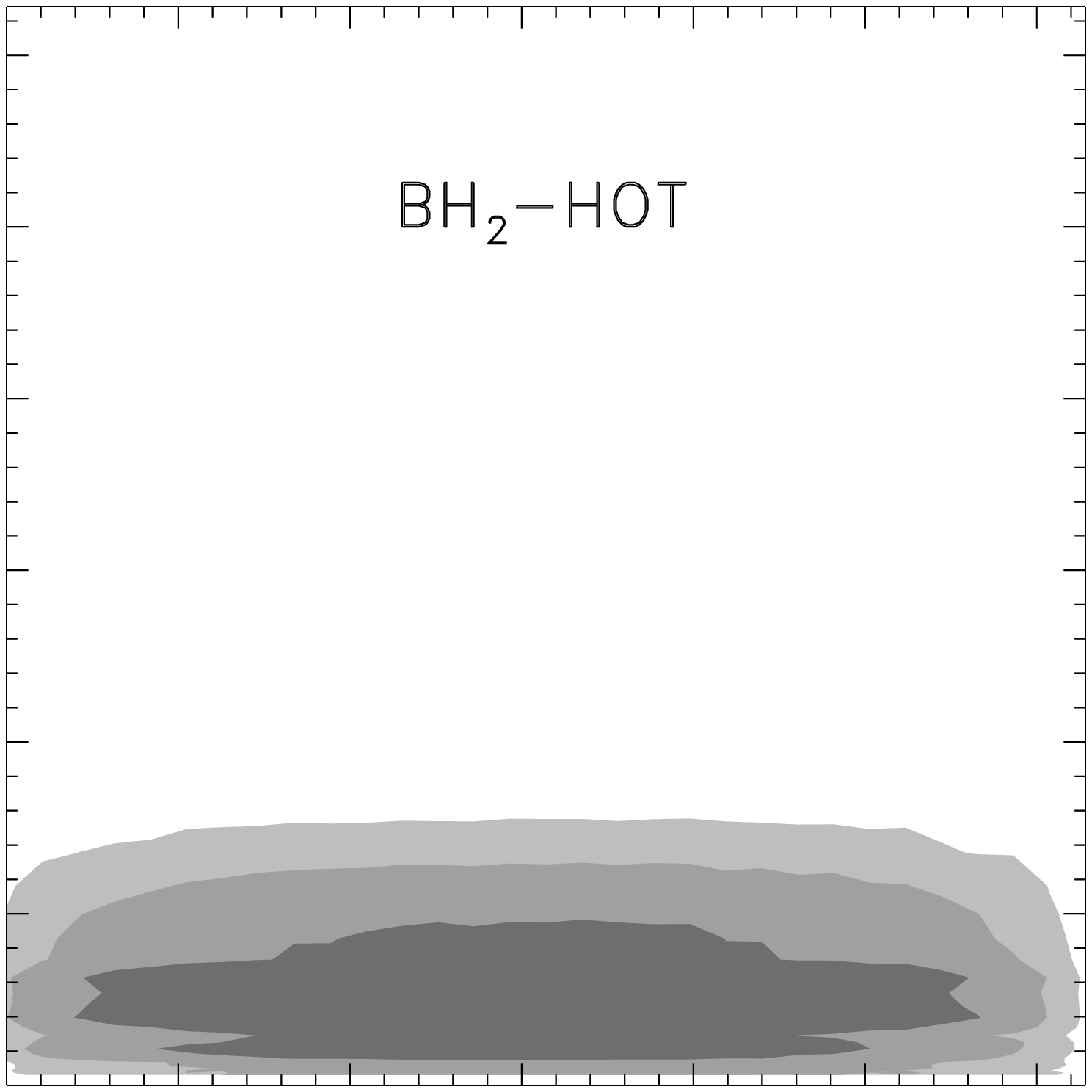}\\
\vskip 1pt
\includegraphics[width=0.48\textwidth]{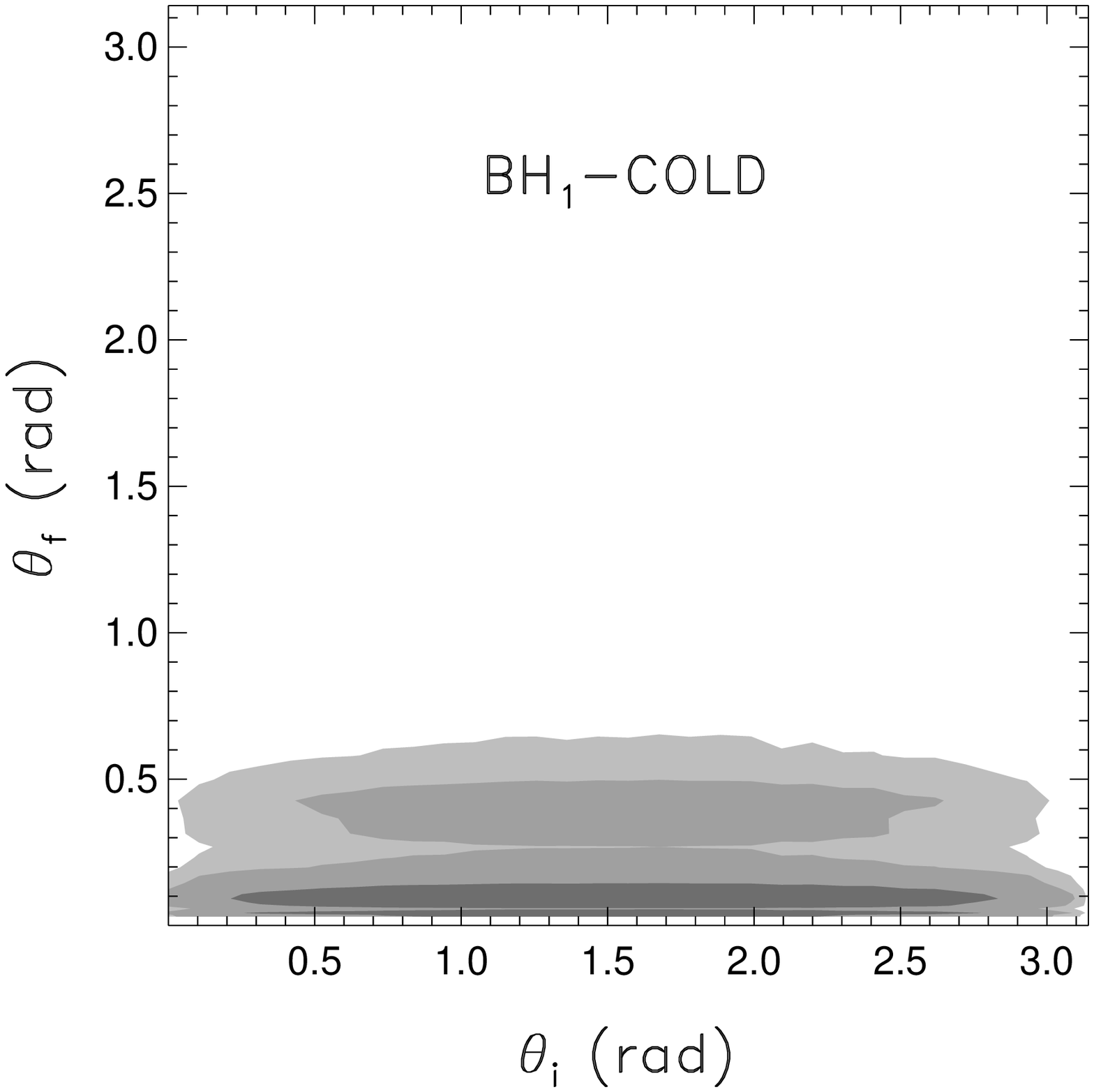}
\includegraphics[width=0.48\textwidth]{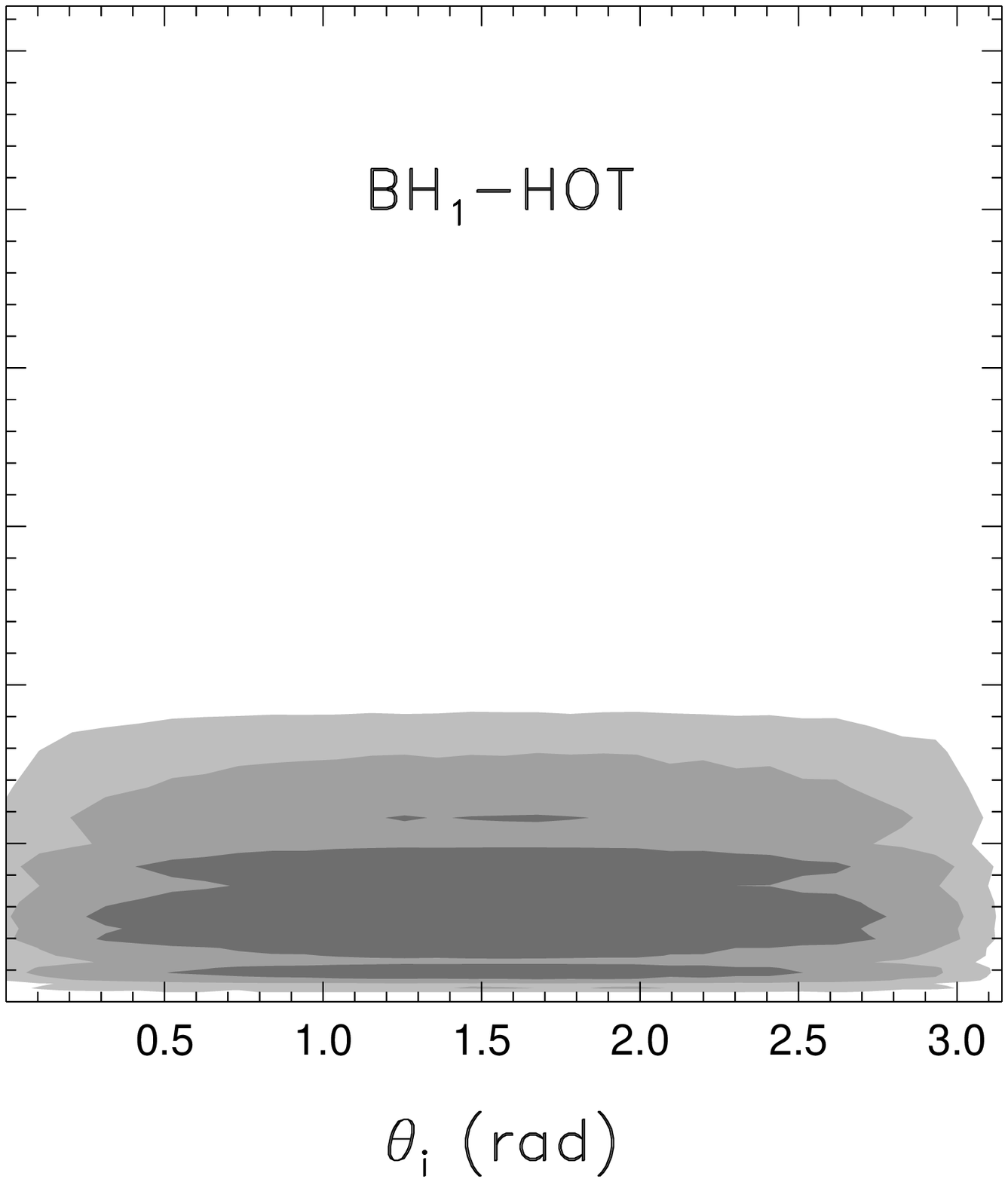}\\

\caption{Density distribution of pairs ($\theta_{\rm i};\theta_{\rm f}$) of the
initial/final relative angles between MBH spins and orbital angular
momentum. Left and right panels refer to MBHs embedded in cold and hot
discs, respectively. Upper (lower) panels refer to the spin of M$_2$
(M$_1$). Dark, medium, and light grey surfaces refer to high density
regions encompassing $68.3\%$, $95.5\%$, and $99.7\%$ of the
realizations.  $\theta_{\rm i}$ has been sampled isotropically, and the
dimensionless spin parameters ($a$) have been sampled from a constant
probability distribution, between 0 and 1. As discussed in the text,
we average $\theta_{\rm f}$ over all the times after the formation of the
MBH binary.}
\label{fig:theta12}
\end{figure*}

Since the time when a binary forms ($t_{\rm bin}$) is different for
different runs, the time intervals ($\Delta t$) over which we average
$\theta_{\rm f}$ are different. We decided to keep constant $\Delta t$ for
runs with the same polytropic index, but we used different $\Delta t$
for cold and hot runs, in order to maximize the statistic. We chose
$\Delta t=3.5$ Myr for runs PC and RC, and $\Delta t=1$ Myr for runs
PH and RH. Averaging over different times does not affect the main
results discussed above. As a check, we computed $\theta_{\rm f}$ and
$\Delta t$ for the two MBHs in runs PC and RC averaging over only 1
Myr, and for every MBH/cold run combination we find $\theta_{\rm f} < 0.19
\, (10^{\circ})$ and $\Delta L_z / \Delta L > 99\%$, consistent with
the values reported in Table~2 for $\Delta t=3.5$ Myr.

We also note that physical processes not implemented in these
simulations, such as star formation or feedback from supernovae, could
decrease the degree of coherency of the accreting gas, possibly
resulting in higher $\theta_{\rm f}$. Furthermore, Lodato et al.  2009
have shown that star formation depletes the reservoir of gas in the
vicinity of the MBHs, and can slow down the decay of the binary at
sub--parsec separations.  A detailed study of the interaction between
star formation in the circumnuclear disc and the properties of the
accreting gas is postponed to a future investigation.

We estimated the efficiency of the alignment process over a large
Monte Carlo sample of initial $\mathbf J_{\rm BH}$. For each MBH and
each run, we selected 20,000 different initial values of $a$,
homogeneously distributed between 0 and 1. For each value of $a$, we
computed the three components of $\mathbf J_{\rm BH}$, assuming an
initially isotropic distribution of the spins. We evolved the initial
condition for $\mathbf J_{\rm BH}$ using the outputs of our
simulations, as described in Section~2.2. 
As already discussed in Section~3, the degree of alignment between MBH
spins and ${\mathbf L}_{\rm pair}$ at $t>t_{\rm bin}$ is ultimately
determined by the gas thermodynamics.  As a consequence, we do not
further analyse the dependence on the initial dynamics of $M_2$, and
focus mainly on the effect of the disc thermodynamics.  The results
from runs CP and CR have been combined in a single class (left panels
in Figure~\ref{fig:theta12}). The same has been done for runs HP and
HR (right panels in Figure~\ref{fig:theta12}).

Figure~\ref{fig:theta12} shows the density of realizations obtained
with our statistical analysis, in the ($\theta_{\rm i}$;$\theta_{\rm
f}$) plane.  Dark, medium, and light grey surfaces refer to regions of
decreasing density, encompassing $68.3\%$, $95.5\%$, and $99.7\%$ of
the realizations. We note that the alignment process is efficient
independently of $\theta_{\rm i}$. The lower density for $\theta_{\rm
i}\approx 0$ and $\theta_{\rm i}\approx \pi$ is due to the initial
isotropic distribution of the spins, and is totally unrelated to the
alignment process.  As already discussed above, alignment is more
efficient for MBHs in cold discs. In these runs (left panels of
Figure~\ref{fig:theta12}) $68.3\%$ of the realizations have a final
angle between the two MBHs and the orbital angular momentum of the
pair $\theta_{\rm f} \lsim 0.1 \, (6^{\circ})$ while $68.3\%$ of the
realizations in runs HP and HR have $\theta_{\rm f} \lsim 0.5 \,
(29^{\circ})$. There are a few $\%$ of the realizations with ``large''
final angles ($\theta_{\rm f} \gsim 0.5 \, (29^{\circ})$) in every
run.

\subsection{Recoil distributions}

In this Section we assume that the two MBHs can reach coalescence, and
we use the distributions of $\theta_{\rm f}$ for $M_1$ and $M_2$ shown
in Figure~\ref{fig:theta12} in order to compute distributions of recoil
velocities for the MBH remnant. We assume also that the distributions
of $\theta_{\rm f}$ we obtained are representative of the relative
angle between the MBH spins and $\mathbf L_{\rm pair}$ during last
phase of orbital decay, when the two MBHs lose efficiently orbital
energy and angular momentum due to the gravitational wave
emission. These assumptions are necessary because our simulations
(spatial resolution $\approx 0.1$ pc) can not follow the evolution of
the MBHs down to separations where gravitational waves dominate the
dynamics.  The two assumptions are valid if one of the following
requirements is fulfilled:

\noindent $\bullet$ The gas accretes onto the two MBHs in a coherent way until star
formation and/or AGN feedback deplete the galactic nucleus of gas, and
no further accretion events (i.e. due to tidal stripping of
stars) change significantly the direction of the MBH spins;

\noindent $\bullet$ The dynamical interaction between the binary and the gas creates
a low density region (the so called ``gap'', Gould \& Rix
2000), reducing/halting accretion onto the MBHs (Milosavljevic \&
Phinney 2005; Dotti et al. 2006b; Hayasaki, Mineshige, Sudou 2007;
Cuadra et al. 2009) so that the spins of the two MBHs do not change
significantly when the binary separation is $\lsim 0.1$ pc;

\noindent $\bullet$ After forming a binary, the two MBH can reach the final
coalescence in a short time ($\lsim 10$ Myr), so that further
accretion events do not have time to change significantly the MBH spin
orientations.

Numerical general relativistic computations show that the recoil
velocity $\mathbf{V}_{\rm kick}$ depends on the binary mass ratio
$q=M_2/M_1$, on the dimensionless spin vectors of the pair
$\mathbf{a}_1$ and $\mathbf{a}_2$ ($0<a_i<1$), and on the orbital
parameters. This information can be obtained from the analysis of
our simulations. We use four different prescriptions from the
literature to compute the recoil velocity of the MBH remnant, based on
Campanelli et al. (2007) and Lousto \& Zlochower (2009; fit CL), Baker
et al. (2008; fit B), Herrmann et al. (2007; fit H), and Rezzolla et
al. (2008; fit R).

We use fit R as a consistency check, as this formula provides the recoil
velocity for completely aligned configurations ($\theta_{\rm f}$ from
the simulations are close to, but not exactly zero), yielding lower
limits for $V_{\rm kick}$.  When using fit R we adopt the spin
magnitudes obtained from our simulations, further assuming that MBH
spins are fully aligned with $\mathbf L_{\rm pair}$ at coalescence,
and $q=1$\footnote{The MBH mass ratio in our simulations is
always between 0.9 and 1. Assuming $q=1$ does not 
affect our results.}.  Expressions of the fitting formulae are
detailed in the Appendix.

\begin{figure*}
\centerline{\psfig{figure=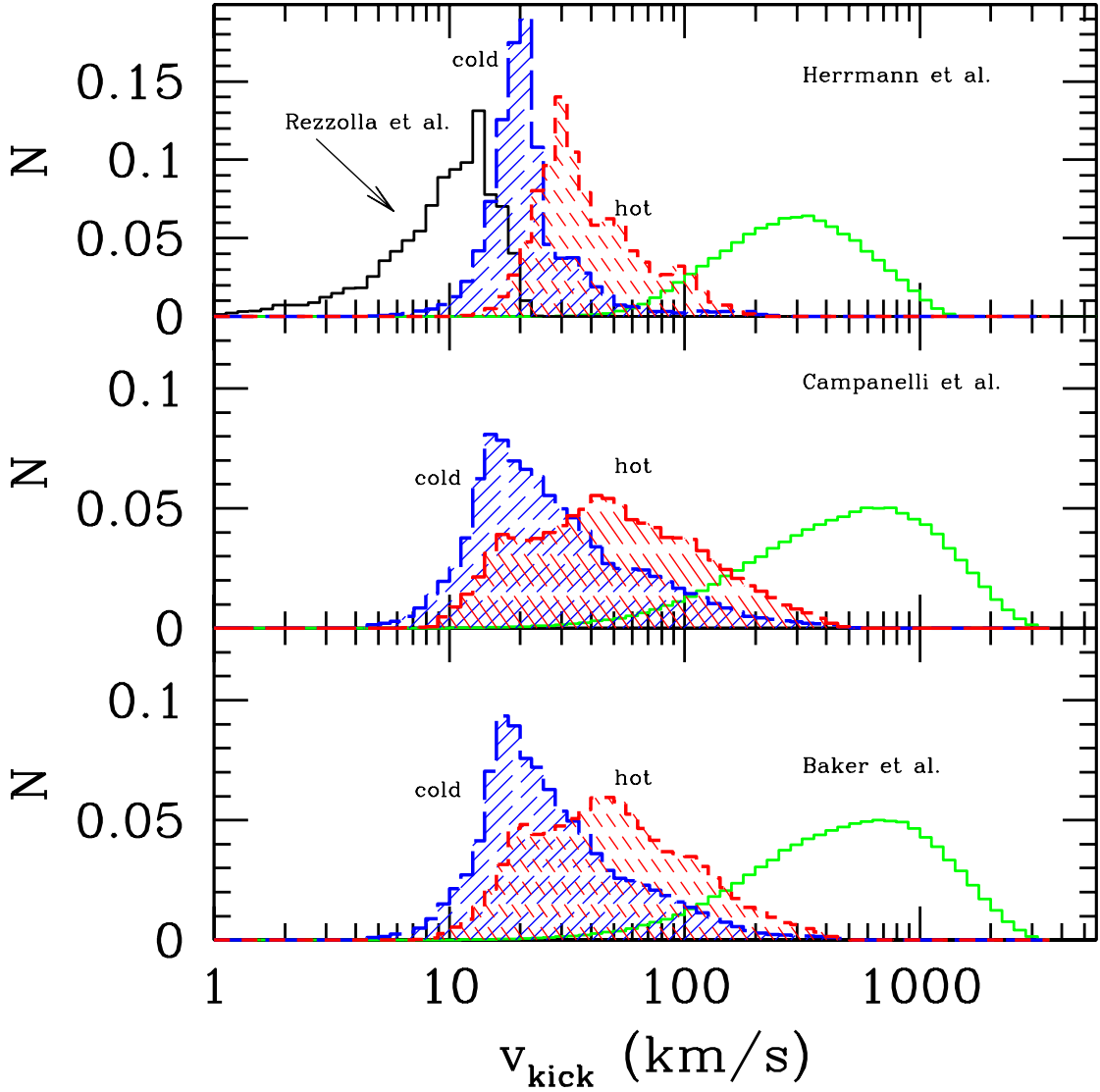,height=16cm}}
\caption{ Distribution of recoil velocities. The blue (red) histogram
is computed from the distribution of spins we obtain from our
simulations, after the alignment of the spins in a cold (hot)
circumnuclear disc.  The green curves refer to recoil velocities
obtained assuming the spins of the two MBHs to be isotropically
distributed.  Upper, middle, and lower panels refer to the results
obtained using fit H, fit CL, and fit B, respectively.  In the upper
panel the black histogram shows as a comparison the distribution of
recoil velocity obtained using fit R, assuming complete alignment
between the MBH spins and ${\mathbf L}_{\rm pair}$. In this case both
the results of cold and hot runs have been considered in a single
histogram. In all the histograms the mass ratio between the MBHs ($q$)
is obtained from our SPH simulations.  }
\label{fig:kick}
\end{figure*}

The distributions of recoil velocities we obtain for cold (blue
histograms) and hot (red histograms) discs are shown in
Figure~\ref{fig:kick}.  For these three fitting formulae we report the
distribution of recoil velocities we would obtain assuming that the
MBH spins are isotropically distributed (green lines).  Because of the
spin alignment discussed in Sections~3.2 and 3.3, the recoil
velocities we obtain analysing the results of our simulations are
approximately one order of magnitude smaller than those predicted for
isotropically distributed MBH spins, independently of the fitting
formula we consider.  Furthermore, the recoils obtained evolving the
MBH pair in a cold circumnuclear disc are always a factor of $\approx
2$ smaller than the velocities obtained in the hot cases.  This shift
is due to the lower level of alignment between MBH spins and $\mathbf
L_{\rm pair}$ for MBHs orbiting in hot discs, as discussed in
Section~3.2.  The mean values of recoil velocities for these three
fitting formulae and for different gas thermodynamics are shown the
first two columns of Table~\ref{tab:kicks}.  Because the mean values
can be affected by the long tails of the recoil distributions at high
velocities, we report also the median values in last two columns of
the same table.

\begin{table}
\begin{center}
\caption{Recoil statistics. All velocities are in $\kms$}
\begin{tabular}{l@{   }c@{   }c@{    }c@{    }c@{    }}  
\hline
 & ~~~cold disc & ~~~hot disc & ~~~cold disc & ~~~hot disc \\
\hline
 & ~~~mean  & ~~~mean  & ~~~median  & ~~~median  \\
\hline
\hline 
\\
Fit CL & 40 $^{+50}_{-40}$ & 62 $^{+42}_{-42}$ & 23 $^{+13}_{-13}$ & 51 $^{+32}_{-32}$ \\
\\
Fit B & 39 $^{+45}_{-39}$ & 67 $^{+61}_{-61}$ & 24 $^{+12}_{-12}$ & 46 $^{+28}_{-28}$ \\
\\
Fit H & 33 $^{+27}_{-27}$ & 54 $^{+33}_{-33}$ & 23 $^{+7}_{-7}$ & 41 $^{+16}_{-16}$ \\
\\
\hline
\end{tabular}\\
\label{tab:kicks}
\end{center}
\noindent
\end{table}

Fit CL, fit B, and fit H give similar mean and median values,
consistent within a factor of $\approx 1 - 1.3$.
The fraction of remnants with recoils larger than in cold (hot) runs
with recoils larger than 400 $\kms$ (``fast recoils'') is 0.2\% (8\%)
using fit CL. Fit B has the same fraction of fast recoils in cold
runs, and a lower fraction (0.2\%) in cold runs. Fit H does not have
any realization with such high recoils.

The black histogram in the upper panel shows the distributions of
recoil velocities obtained using fit R. In this case we considered
both the results of cold and hot runs in a single histogram. Mean and
median values for the recoils are $\approx 10 \kms$.  Such low values
follow from using the distributions of $a_i$ that we obtain from our
simulations. After the formation of a binary, the MBHs in our runs
have spin magnitudes $0.3 \lsim a_i \lsim 0.9$.  If instead we assumed
a homogeneous distribution of spins between 0 and 1, fit R would
predict a kick distribution with a peak at $\approx 100 \kms$, a sharp
cutoff at higher velocities, and a long tail at lower values.  As
expected, fit R gives a lower limit for the recoil velocities.  The
recoil distribution obtained with this last prescription peaks at
velocities which are only $\approx 2$ times smaller than those where
the cold runs peak, when using the other three fits. This confirms
that these fitting formulae well describe quasi--aligned
configurations.

\section{Discussion}

In this paper, we traced for the first time the evolution of the spin
vectors of MBHs orbiting inside a massive circumnuclear gas disc. Our SPH
simulations have sufficiently high resolution to probe the
hydrodynamics of fluid particles and the accretion physics near the
gravitational sphere of influence of the MBHs. An ad-hoc algorithm
designed for tracking the gravo--magnetic coupling between the MBH spin
and the small-scale accretion disc is then implemented in the code.
We find that:

\noindent $\bullet$ When evolving a in dense, rotationally supported,
structure such as a circumnuclear disc, MBHs in a pair align their
spins ($\mathbf J_{\rm BH_{1,2}}$) to the pair orbital angular
momentum ($\mathbf L_{\rm pair}$) well before the two MBHs bind in a
Keplerian binary, and independently of the MBHs initial orbital
parameters.  For a run with $M_2$ initially on a retrograde orbit, the
spin of the secondary aligns efficiently only after the ``orbital
angular momentum flip''.

\noindent $\bullet$ The average angle between $\mathbf J_{\rm BH_{1,2}}$ and
$\mathbf L_{\rm pair}$ after the binary formation ($\theta_{\rm f}$)
depends on the thermodynamics of the massive circumnuclear discs.
$\theta_{\rm f}$ is lower if the MBHs are embedded in colder discs
(with a polytropic index $\gamma=7/5$), with respect to hotter discs
($\gamma=5/3$);

\noindent $\bullet$ After the formation of a binary, the two MBHs accrete
  gas with the same dynamical and thermodynamical properties. As a
  consequence, even the angle between the two small projections of
  $\mathbf J_{\rm BH_{1,2}}$ in the orbital plane decreases. This
  further reduces the recoil velocity of the MBH remnant. The degree
  of alignment between the two spins and between each spin and
  $\mathbf L_{\rm pair}$ is preserved (or even increased) by
  spin--spin and spin--orbit interactions until the plunge phase
  (Schnittman 2004; Herrmann et al.  2009);

\noindent $\bullet$ Due to the efficient alignment between $\mathbf J_{\rm
  BH_{1,2}}$ and $\mathbf L_{\rm pair}$, the expected recoil
velocities ($V_{\rm kick}$) at the MBH coalescence is, on average, one
order of magnitude lower than those expected for randomly oriented MBH
spins.  The thermodynamical properties of the environment affect the
degree of alignment and, as a consequence, the expected recoil
velocities. $V_{\rm kick}$ is lower (by a factor of 1.5--2.2) for
lower values of $\gamma$;

\noindent $\bullet$  Assuming the same distribution of
$\theta_{\rm f}$, the recoil velocity distributions obtained using
different prescriptions are very similar. The three fitting formulae
used predict the same mean and median recoil velocities ($\lsim 70
\kms$) within a factor of $\approx 1 - 1.3$, with less than few
percent of realizations having $V_{\rm kick}>400 \kms$.

The distributions of recoil velocities that we find have important
consequences for retention of MBHs in galactic nuclei.  When MBH
binaries form and evolve in gas--rich major mergers, we predict the
recoil velocity to be, on average, well below the escape speed from
low-redshift galaxies. Indeed, because of the extreme efficiency of
the spin alignment process, the recoil velocities are likely
unimportant even for high-z proto-galactic building blocks.  Volonteri
\& Rees (2006) and Volonteri (2007) discussed how strong recoils can
affect the early growth of MBHs at the highest redshifts.  In a
forthcoming paper we will update our calculations and determine the
impact of low recoils on the growth of MBHs in galaxies.

Our simple treatment of thermodynamics and the absence of any
prescription for star--formation and supernovae feedback in our
simulations could, in principle, overestimate the degree of coherency
in the gas flows accreting onto the MBHs. Furthermore, our finite
resolution prevent us to study the fragmentation of the accretion
discs forming around the MBHs, that could result in a sequence of
short and randomly oriented accretion events (King \& Pringle 2006).
We will investigate interaction between star formation in the
circumnuclear disc and the properties of the accreting gas in a
forthcoming study.

\section*{Acknowledgments}

The authors thank the anonymous Referee for her/his suggestions that
have improved the scientific content of the paper. We are grateful to
Kayhan Gultekin for fruitful discussions and technical help, and Jon
Gair for valuable suggestion and coffee-support.  We also thank John
Baker, Emanuele Berti, Manuela Campanelli, Pablo Laguna, Cole Miller,
Denis Pollney, Luciano Rezzolla, and James van Meter for comments and
kind clarifications on the results of numerical relativity
simulations.  MD thanks Luca Paredi for the technical support. MV
acknowledges support from a Rackham faculty grant.

\section*{Appendix A: Fitting formulae for recoil velocities}

Campanelli et al. (2007) and Lousto \& Zlochower (2009; fit CL) propose the
following fitting formula for the post--coalescence recoil of a MBH
remnant:

\begin{eqnarray}
{V}_{\rm kick} &=& \sqrt{v_m^2 + v_{\perp}^2+2 v_m v_{\perp} \cos(\xi)+ v_{\parallel}^2},
      \label{eq:v_total}\\ v_m &=& A \eta^2 \sqrt{1 - 4 \eta}\, (1 + B
      \eta), \label{eq:v_mass}\\ v_{\perp} &=& \frac {H
      \eta^2}{(1+q)}\left( a_1^{\parallel} - q a_2^{\parallel}
      \right), \label{eq:v_perp}\\ v_{\parallel} &=& \frac{K
      \eta^2}{(1+q)}\,\cos(\Theta-\Theta_0)|{\mathbf a}_1^{\perp} - q
      {\mathbf a}_2^{\perp}|, \label{eq:v_parallel}
\end{eqnarray}
where $A = 1.2 \times 10^4 \kms$, $B = -0.93$, $H = 6900\kms$, $K =
6.0 \times 10^4 \kms$, $\eta\equiv q/(1+q)^2$ is the symmetric mass
ratio and $\xi$ measures the angle between the unequal mass and the
spin contribution to the recoil velocity in the orbital plane. We
assumed $\xi=145^{\circ}$, as suggested by Lousto \& Zlochower.  The
components of the spins of the two MBHs are:

\begin{eqnarray*}
              a_1^{\perp}&=&a_1\, \sin(\theta_1)\\
              a_1^{\parallel}&=&a_1\, \cos(\theta_1)\\
              a_2^{\perp}&=&a_2\, \sin(\theta_2)\\
              a_2^{\parallel}&=&a_2\, \cos(\theta_2),\\
\end{eqnarray*}
where the indices ${\parallel}$ and ${\perp}$ refer to projections
parallel and perpendicular to the orbital angular momentum,
respectively, and $\theta_1$ ($\theta_2$) refers to $\theta_{\rm f}$
for the primary (secondary) MBH.  In eq.~7, $\Theta$ is the angle
between $({\mathbf a}_2^{\perp} - q {\mathbf a}_1^{\perp})$ and the
separation vector at coalescence, and $\Theta_0$ depends on the
initial separation between the holes. Since $\Theta_0$ is unknown, for
this exploration we assume a flat distribution of $\Theta-\Theta_0$
between 0 and $2\pi$.

Baker et al. (2008; fit B) propose instead the following fitting formula for
$v_{\parallel}$ :
\begin{eqnarray}
v_{\parallel} &=& \frac{K \eta^3}{(1+q)}\,\left(a_1^{\perp} \cos(\phi_1-\Phi_1)- q a_2^{\perp} \cos(\phi_2-\Phi_2)\ \right),
\end{eqnarray}
where $\phi_1$ ($\phi_2$) is the angle between $a_{1}^{\perp}$
($a_{2}^{\perp}$) and a fixed reference direction. Following Baker et
al. (2008), $\Phi_1=\Phi(q)$ and $\Phi_2=\Phi(1/q)$. Because in our
simulations $q \approx 1$, we fixed $\Phi_1=\Phi_2$. We further assume a
flat distribution of $\Phi_1$ between 0 and $2\pi$. In this case, $A =
1.35 \times 10^4 \kms$, $B = -1.48$, $H = 7540 \kms$, and $K = 2.4
\times 10^5 \kms$.

Herrmann et al. (2007; fit H) formulate the recoil velocity of a MBH
remnant as a function of a different angle $\theta_{\rm H}$, i.e.  the
angle between $\mathbf L_{\rm pair}$ and
\begin{equation}
{\mathbf \Sigma}=M\left(\frac{{\mathbf J}_2}{M_2} - \frac{{\mathbf J}_1}{M_1} \right),
\end{equation}
where $M=M_1+M_2$. Assuming that $\mathbf L_{\rm pair}$ is aligned
with the $z$ direction, they find that the Cartesian component of the
recoil velocity follow:
\begin{eqnarray}
V_x &=& C_0 H_x \cos(\theta_{\rm H}),   \\
V_y &=& C_0 H_y \cos(\theta_{\rm H}),   \\
V_z &=& C_0 K_z \sin(\theta_{\rm H}),   
\end{eqnarray}
where $C_0=\Sigma q^2 / (M^2 (1+q)^4)$, and the best fitting
parameters are $H_x= 2.1\times 10^3$ $H_y=7.3\times 10^3 $, and
$K_z=2.1\times 10^4$\footnote{The values of $H_x$, $H_y$, and $K_z$
published in Herrmann et al. (2007) contain a typo (Laguna, private
communication).}.

 The last fitting formula we use to compute $V_{\rm kick}$ has been
proposed by Rezzolla et al. (2008; fit R). In their study they
consider only equal--mass MBHs with spins aligned with $\mathbf L_{\rm
pair}$. They find:
\begin{equation}
{V}_{\rm kick}=|c_1 (a_1 -a_2) + c_2 (a_1^2-a_2^2)|,
\label{eq:pollney}
\end{equation}
where $c_1=-220.97$ and $c_2=45.52$.  Eq.~\ref{eq:pollney} provides
recoil velocities for completely aligned configurations.

\end{document}